\documentclass[prd,aps,preprintnumbers,floatfix,showpacs,preprint,tightenlines]{revtex4}
\usepackage{epsfig}
\usepackage{graphics}
\usepackage{latexsym}
\usepackage{amsmath}
\usepackage{amssymb}
\usepackage{rotating}

\begin{document}
\preprint{ \vbox{ \hbox{JLAB-THY-06-593}  \hbox{ADP-07-01/T641} }}

\title{Chiral extrapolation of nucleon magnetic form factors}

\author{P. Wang$^{abc}$}
\author{D. B. Leinweber$^b$}
\author{A. W. Thomas$^{c}$}
\author{R. D. Young$^c$}

\affiliation{ $^a$Physics Department, North Carolina State University,
Raleigh, NC 27695, USA}

\affiliation{ $^b$Special Research Center for the Subatomic Structure
of Matter (CSSM) and Department of Physics, University of Adelaide
5005, Australia}

\affiliation{ $^c$Jefferson Laboratory, 12000 Jefferson Ave., Newport
News, VA 23606 USA}

\begin{abstract}
The extrapolation of nucleon magnetic form factors calculated within
lattice QCD is investigated within a framework based upon heavy baryon
chiral effective-field theory.  All one-loop graphs are considered at
arbitrary momentum transfer and all octet and decuplet baryons are
included in the intermediate states.  Finite range regularisation is
applied to improve the convergence in the quark-mass expansion.  At each value
of the momentum transfer ($Q^2$), a separate extrapolation to the
physical pion mass is carried out as a function of $m_\pi$ alone.
Because of the large values of $Q^2$ involved, the role of the pion
form factor in the standard pion-loop integrals is also investigated.
The resulting values of the form factors at the physical pion mass are
compared with experimental data as a function of $Q^2$ and demonstrate
the utility and accuracy of the chiral extrapolation methods
presented herein.
\end{abstract}


\maketitle

\section{Introduction}

The study of the electromagnetic properties of the nucleon is of
great importance in understanding the structure of baryons --- see
Refs.~\cite{Gao:2003ag,Hyde-Wright:2004gh,Arrington:2006zm,Perdrisat:2006hj}
for recent reviews. The most rigorous approach to low-energy
phenomena in QCD is via numerical simulations in lattice gauge
theory and many physical quantities, such as baryon masses, magnetic
moments, etc.\
\cite{Leinweber1,Leinweber2,Leinweber0,Leinwebernew,Edwards:2005ym,Boinepalli:2006xd,QCDSF}
have been investigated within lattice QCD. Because of computing
limitations, most of those quantities are simulated with large quark
($\pi$) masses and an extrapolation of lattice results to the
physical $\pi$ mass is needed. Early lattice extrapolations
considered simple polynomial functions of $\pi$ mass. However, it is
now widely acknowledged that the chiral non-analytic behavior
predicted by chiral perturbation theory ($\chi$PT) must be
incorporated in any quark mass extrapolation function
\cite{Leinweber:1993hj,Leinweber:2001ui,Hemmert:2002uh,Leinweber:1999ig,Detmold:2001jb}.

$\chi$PT has been a very useful approach to the study of low
momentum processes involving mesons and baryons and has been used in
various studies of baryon structure. It is based on an
effective Lagrangian constructed in a systematic way and consistent
with all the symmetries of QCD. The first systematic discussion for
the two flavor sector, i.e. the pion-nucleon system, of how to
implement the ideas of chiral power counting \cite{Weinberg}, was
performed in Ref.~\cite{Gasser}. However, treating the nucleons as
relativistic Dirac fields does not allow for a one-to-one
correspondence between the expansion in small momenta and quark mass
on the one hand and pion loops on the other. As pointed out in
Ref.~\cite{Jenkins}, this shortcoming can be overcome if one makes
use of methods borrowed from heavy quark effective field theory
(HQEFT), namely to consider the baryons as extremely heavy, static
sources. The relativistic or heavy baryon chiral perturbation theory
has been applied to study a range of hadron properties in QCD,
including nucleon magnetic moments and charge radii
\cite{Durand1,Kubis}, the nucleon sigma commutator
\cite{Borasoy:1996bx,Leinweber3,Procura} and moments of structure
functions \cite{Detmold:2001jb,Hemmert}.

Historically, most formulations of $\chi$PT are based on dimensional
or infrared regularisation. However, the physical predictions of
effective field theory must be regularisation scheme independent,
such that other schemes are possible and may provide advantages over
the traditional approach. Indeed, Donoghue $et$ $al$.\
\cite{Donoghue} have already reported the improved convergence of
properties of effective theory formulated with what they called a
``long-distance regulator". With the most detailed studies being on
the extrapolation of the nucleon mass, it has been shown that the
use of finite range regularisation (FRR) enables the most
systematically accurate connection of $\chi$PT and lattice
simulation results
\cite{Leinweber3,Leinweber4,Young2,Armour:2005mk,Allton:2005fb}.

The FRR $\chi$PT has been applied to the extrapolation of proton
magnetic moment with the leading non-analytic contribution of pions
\cite{Young:2004tb}. It was found that the smooth behavior of the lattice
data, together with the series truncations of the FRR expansion
indicate that although higher order terms of DR can be individually
large they effectively sum to zero in the region of interest. FRR
$\chi$PT provides a resummation of the chiral expansion that ensures
that the slow variation of magnetic moments observed in lattice QCD
arises naturally in the FRR expansion. It was also predicted that the
quenched and physical magnetic moments are in good agreement over a
large range of pion mass, especially at large $m_\pi$ \cite{Young:2004tb}.

In this paper, we will extrapolate the proton and neutron magnetic
moments, as well as the form factors at finite momentum transfer,
within a framework based upon heavy baryon chiral perturbation
theory. Many methods have been used to compute the form
factors at the physical value of the pion mass, including the early
relativistic approach \cite{Jenkins}, heavy baryon chiral
perturbation theory \cite{Bernard:1992qa}, the so-called small scale
expansion \cite{Bernard:1998gv}, relativistic chiral perturbation
theory \cite{Kubis,Fuchs:2003ir}, etc.. The spectral functions of
the form factors have also been investigated by calculating the imaginary
parts of the form factors \cite{Bernard:1996cc,Kaiser:2003qp}. The
disappointing observation was that a satisfactory description of the
electromagnetic form factors was achieved only up to $Q^2\sim 0.1$
GeV$^2$. In order to improve this situation it is natural to
consider higher order terms in chiral perturbation theory but eventually
the series must diverge. In fact,
effectively resumming the series by including vector meson degrees
of freedom led to a satisfactory
description of the electromagnetic form factors up to
$Q^2\sim 0.4$ GeV$^2$ \cite{Kubis,Schindler:2005ke}. It is of
interest to calculate the form factors at relatively large momentum
transfer and pion mass because there are many
experimental data in this region and
it is a priority for lattice QCD to understand that data.

Because the values of the momentum transfer are quite large, it is
not possible to make a systematic expansion in both $Q^2$ and
$m_\pi$. Instead, we extrapolate as a function of $m_\pi$ at {\em
each} separate value of $Q^2$. All the one loop contributions,
including baryon octet and decuplet intermediate states, are
considered. The quenched lattice data at large quark mass are used
in the extrapolation, using the finding that the difference between
quenched and full QCD data is usually quite small at large values of
the pion mass. While this is a reasonable approach until high
quality full QCD data is available (for first dynamical studies see
Refs.~\cite{Alexandrou:2006ru,Gockeler:2006ui,Edwards:2006qx}), it
does mean that in comparing with experiment we must remember that
there is an unknown systematic error associated with the use of
quenched data. So too, we have not had lattice data available which
would permit an extrapolation to the continuum ($a \rightarrow 0$)
and infinite volume limits. In spite of all these caveats the
results of this exploratory study are really very promising.

\section{Chiral Perturbation Theory}

There are many papers which deal with heavy baryon chiral perturbation
theory -- for details see, for example, Refs.
\cite{Jenkins2,Labrenz,Durand2,Tiburzi:2004mv}. For completeness, we briefly
introduce the formalism in this section. In the heavy baryon chiral
perturbation theory, the lowest chiral Lagrangian for the baryon-meson
interaction which will be used in the calculation of the nucleon
magnetic moments, including the octet and decuplet baryons, is
expressed as
\begin{eqnarray}
{\cal L}_v &=&i{\rm Tr}\bar{B}_v(v\cdot {\cal D})
B_v+2D{\rm Tr}\bar{B}_v S_v^\mu\{A_\mu,B_v\}
+2F{\rm Tr}\bar{B}_v S_v^\mu[A_\mu,B_v]
\nonumber \\
&& -i\bar{T}_v^\mu(v\cdot {\cal D})T_{v\mu}
+{\cal C}(\bar{T}_v^\mu A_\mu B_v+\bar{B}_v A_\mu T_v^\mu),
\end{eqnarray}
where $S_\mu$ is the covariant spin-operator defined as
\begin{equation}
S_v^\mu=\frac i2\gamma^5\sigma^{\mu\nu}v_\nu.
\end{equation}
Here, $v^\nu$ is the nucleon four velocity (in the rest frame, we have
$v^\nu=(1,0)$).
D, F and $\cal C$ are the coupling constants.
The chiral covariant derivative $D_\mu$ is written as $D_\mu
B_v=\partial_\mu B_v+[V_\mu,B_v]$. The pseudoscalar meson octet
couples to the baryon field through the vector and axial vector
combinations
\begin{equation}
V_\mu=\frac12(\zeta\partial_\mu\zeta^\dag+\zeta^\dag\partial_\mu\zeta),~~~~~~
A_\mu=\frac12(\zeta\partial_\mu\zeta^\dag-\zeta^\dag\partial_\mu\zeta),
\end{equation}
where
\begin{equation}
\zeta=e^{i\phi/f}, ~~~~~~
f=93~{\rm MeV}.
\end{equation}
The matrix of pseudoscalar fields $\phi$ is expressed as
\begin{eqnarray}
\phi=\frac1{\sqrt{2}}\left(
\begin{array}{lcr}
\frac1{\sqrt{2}}\pi^0+\frac1{\sqrt{6}}\eta & \pi^+ & K^+ \\
\pi^- & -\frac1{\sqrt{2}}\pi^0+\frac1{\sqrt{6}}\eta & K^0 \\
K^- & \bar{K}^0 & -\frac2{\sqrt{6}}\eta
\end{array}
\right).
\end{eqnarray}
$B_v$ and $T^\mu_v$ are the velocity dependent new fields
which are related to the original baryon octet and decuplet fields
$B$ and $T^\mu$ by
\begin{equation}
B_v(x)=e^{im_N \not v v_\mu x^\mu} B(x),
\end{equation}
\begin{equation}
T^\mu_v(x)=e^{im_N \not v v_\mu x^\mu} T^\mu(x).
\end{equation}
In the chiral $SU(3)$ limit, the octet baryons will have the same
mass $m_B$. In our calculation, we use the physical masses for
baryon octets and decuplets. The explicit form of the baryon octet
is written as
\begin{eqnarray}
B=\left(
\begin{array}{lcr}
\frac1{\sqrt{2}}\Sigma^0+\frac1{\sqrt{6}}\Lambda &
\Sigma^+ & p \\
\Sigma^- & -\frac1{\sqrt{2}}\Sigma^0+\frac1{\sqrt{6}}\Lambda & n \\
\Xi^- & \Xi^0 & -\frac2{\sqrt{6}}\Lambda
\end{array}
\right).
\end{eqnarray}
For the baryon decuplets, there are three indices, defined as
\begin{eqnarray}
T_{111}=\Delta^{++}, ~~ T_{112}=\frac1{\sqrt{3}}\Delta^+, ~~
T_{122}=\frac1{\sqrt{3}}\Delta^0, \\ \nonumber
T_{222}=\Delta^-, ~~ T_{113}=\frac1{\sqrt{3}}\Sigma^{\ast,+}, ~~
T_{123}=\frac1{\sqrt{6}}\Sigma^{\ast,0}, \\ \nonumber
T_{223}=\frac1{\sqrt{3}}\Sigma^{\ast,-}, ~~
T_{133}=\frac1{\sqrt{3}}\Xi^{\ast,0}, ~~ T_{233}=\frac1{\sqrt{3}}\Xi^{\ast,-},
~~ T_{333}=\Omega^{-}.
\end{eqnarray}

The octet, decuplet and octet-decuplet transition magnetic moment
operators are needed in the one loop calculation of nucleon magnetic
form factors. The baryon octet magnetic Lagrangian is written as:
\begin{equation}\label{lomag}
{\cal L}=\frac{e}{4m_N}\left(\mu_D{\rm Tr}\bar{B}_v \sigma^{\mu\nu}
\left\{F^+_{\mu\nu},B_v\right\}+\mu_F{\rm Tr}\bar{B}_v \sigma^{\mu\nu}
\left[F^+_{\mu\nu},B_v \right]\right),
\end{equation}
where
\begin{equation}
F^+_{\mu\nu}=\frac12\left(\zeta^\dag F_{\mu\nu}Q\zeta+\zeta
F_{\mu\nu}Q\zeta^\dag\right).
\end{equation}
$Q$ is the charge matrix $Q=$diag$\{2/3,-1/3,-1/3\}$. At the lowest
order, the Lagrangian will generate the following nucleon magnetic
moments:
\begin{equation}\label{treemag}
\mu_p=\frac13\mu_D+\mu_F,~~~~~~ \mu_n=-\frac23\mu_D.
\end{equation}

The decuplet magnetic moment operator is expressed as
\begin{equation}
{\cal L}=-i\frac{e}{m_N}\mu_C q_{ijk}\bar{T}^\mu_{v,ikl}T^\nu_{v,jkl}
F_{\mu\nu},
\end{equation}
where $q_{ijk}$ and $q_{ijk}\mu_C$ are the charge and magnetic
moment of the decuplet baryon $T_{ijk}$.
The transition magnetic operator is
\begin{equation}
{\cal L}=i\frac{e}{2m_N}\mu_T F_{\mu\nu}\left(\epsilon_{ijk}Q^i_l \bar{B}^j_{vm} S^\mu_v
T^{\nu,klm}_v+\epsilon^{ijk}Q^l_i \bar{T}^\mu_{v,klm} S^\nu_v B^m_{vj}\right).
\end{equation}
In Ref.~\cite{Durand1}, the authors used $\mu_u$, $\mu_d$ and
$\mu_s$ instead of the $\mu_C$ and $\mu_T$. For the particular
choice, $\mu_s=\mu_d=-\frac12 \mu_u$, one finds the following
relationship:
\begin{equation}
\mu_D=\frac32 \mu_u, ~~~ \mu_F=\frac23 \mu_D, ~~~ \mu_C=\mu_D, ~~~ \mu_T=-4\mu_D.
\end{equation}
In our numerical calculations, the above formulas are used and
therefore all baryon magnetic moments are related to one parameter,
$\mu_D$.

In the heavy baryon formalism, the propagators of the octet or
decuplet baryon, $j$, are expressed as
\begin{equation}
\frac i {v\cdot k-\delta^{jN}+i\varepsilon} ~~{\rm and}~~ \frac {iP^{\mu\nu}}
{v\cdot k-\delta^{jN}+i\varepsilon},
\end{equation}
where $P^{\mu\nu}$ is $v^\mu v^\nu-g^{\mu\nu}-(4/3)S_v^\mu S_v^\nu$.
$\delta^{ab}=m_b-m_a$ is the mass difference of between the two
baryons. The propagator of meson $j$ ($j=\pi$, $K$, $\eta$) is the usual
free propagator, i.e.:
\begin{equation}
\frac i {k^2-M_j^2+i\varepsilon}.
\end{equation}

\begin{center}
\begin{figure}[tbp]
\includegraphics[scale=0.9]{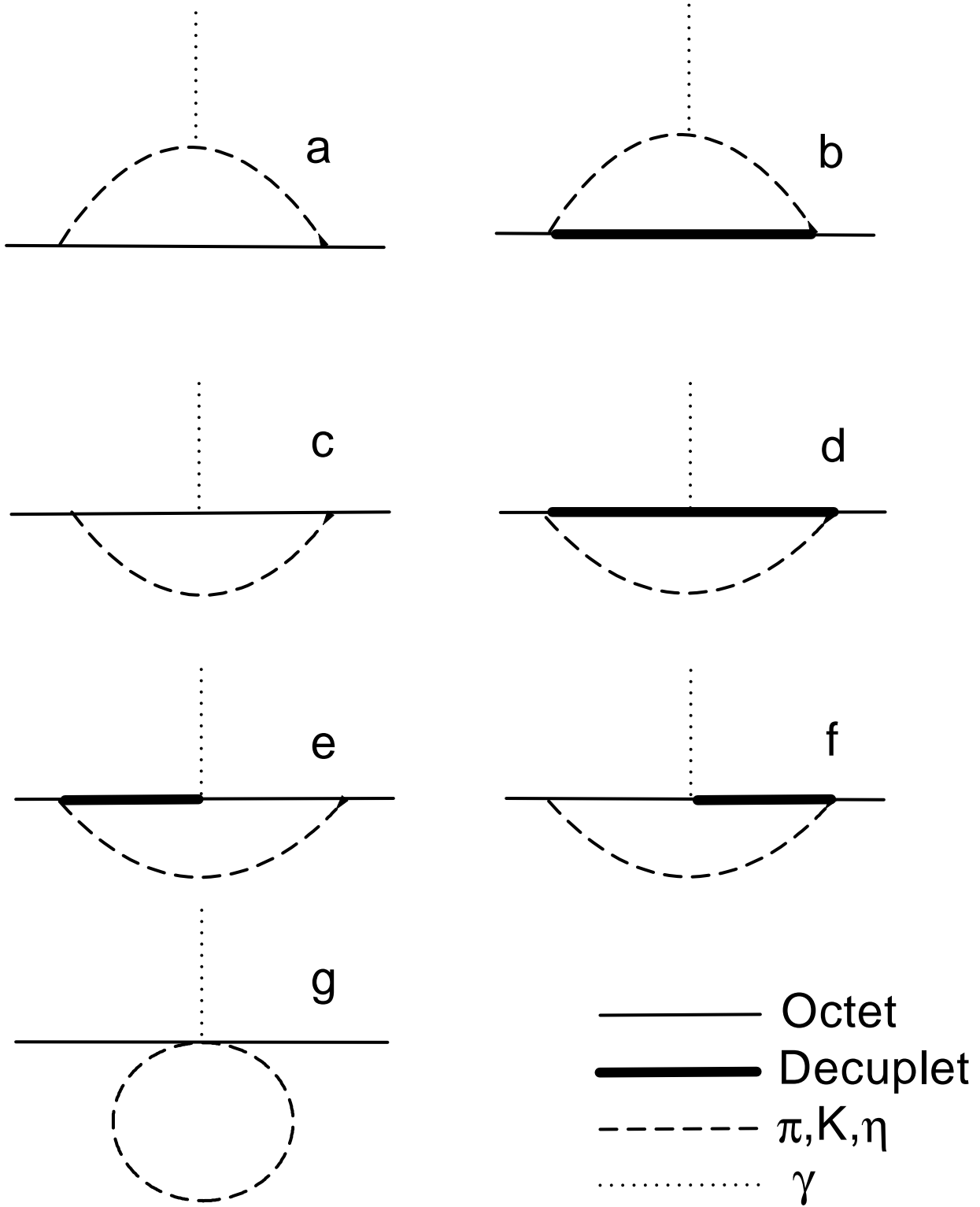}
\caption{The one loop Feynman diagrams for the nucleon magnetic moments.
The solid, thick solid, dash and dotted lines are for the octet baryons,
decuplet baryons, pseudoscalar mesons, and photons, respectively.}
\end{figure}
\end{center}

\section{Nucleon Magnetic Moments}

In the heavy baryon formalism, the nucleon form factors are
defined as:
\begin{equation}
<B(p^\prime)|J_\mu|B(p)>=\bar{u}(p^\prime)\left\{v_\mu
G_E(Q^2)+\frac{i\epsilon_{\mu\nu\alpha\beta}v^\alpha S_v^\beta
q^\nu}{m_N}G_M(Q^2)\right\}u(p),
\end{equation}
where $q=p^\prime-p$ and $Q^2=-q^2$. According to the Lagrangian,
the one loop Feynman diagrams which contribute to the nucleon
magnetic moments are plotted in Fig.~1. The contributions to nucleon
magnetic form factors of Fig.~1a are expressed as
\begin{equation}\label{p1a}
G_M^{p(1a)}=\frac{m_N(D+F)^2}{8\pi^3f_\pi^2}I_{1\pi}^{NN}
+\frac{m_N(D+3F)^2I_{1K}^{N\Lambda}+3m_N(D-F)^2I_{1K}^{N\Sigma}}{{48\pi^3f_\pi^2}},
\end{equation}
\begin{equation}\label{n1a}
G_M^{n(1a)}=-\frac{m_N(D+F)^2}{8\pi^3f_\pi^2}I_{1\pi}^{NN}
+\frac{m_N(D-F)^2}{8\pi^3f_\pi^2}I_{1K}^{N\Sigma}.
\end{equation}
The integration $I_{1j}^{\alpha\beta}$ is expressed as
\begin{equation}
I_{1j}^{\alpha\beta}=\int d\overrightarrow{k}\frac{k_y^2
u(\overrightarrow{k}+\overrightarrow{q}/2)
u(\overrightarrow{k}-\overrightarrow{q}/2)(\omega_j(\overrightarrow{k}+\overrightarrow{q}/2)
+\omega_j(\overrightarrow{k}-\overrightarrow{q}/2)+\delta^{\alpha\beta})}
{A_j^{\alpha\beta}},
\end{equation}
where
\begin{eqnarray}
A_j^{\alpha\beta}&=&\omega_j(\overrightarrow{k}+\overrightarrow{q}/2)
\omega_j(\overrightarrow{k}-\overrightarrow{q}/2)
(\omega_j(\overrightarrow{k}+\overrightarrow{q}/2)+\delta^{\alpha\beta})
\nonumber \\
&&(\omega_j(\overrightarrow{k}-\overrightarrow{q}/2)+\delta^{\alpha\beta})
(\omega_j(\overrightarrow{k}+\overrightarrow{q}/2)+\omega_j(\overrightarrow{k}-\overrightarrow{q}/2)).
\end{eqnarray}
$\omega_j(\overrightarrow{k})=\sqrt{m_j^2+\overrightarrow{k}^2}$ is
the energy of the meson $j$. In our calculation we use the finite
range regularisation and $u(\overrightarrow{k})$ is the ultra-violet
regulator. This diagram is studied in the previous paper \cite{Young2}
which gives the leading analytic term to the magnetic moments.  The
first terms in Eqs.\ (\ref{p1a}) and (\ref{n1a}) come from the $\pi$
meson cloud contribution. The second terms come from the K meson cloud
contribution. Fig.~1b is the same as Fig.~1a but the intermediate
states are decuplet baryons.  Their contributions to the magnetic form
factors are expressed as
\begin{equation}
G_M^{p(1b)}=\frac{m_N{\cal C}^2}{36\pi^3f_\pi^2}I_{1\pi}^{N\Delta}
-\frac{m_N{\cal C}^2}{144\pi^3f_\pi^2}I_{1K}^{N\Sigma^\ast},
\end{equation}
\begin{equation}
G_M^{n(1b)}=-\frac{m_N{\cal C}^2}{36\pi^3f_\pi^2}I_{1\pi}^{N\Delta}
-\frac{m_N{\cal C}^2}{72\pi^3f_\pi^2}I_{1K}^{N\Sigma^\ast}.
\end{equation}
The contributions to the form factors from Fig.~1c are expressed as
\begin{eqnarray} \nonumber
G_M^{p(1c)}&=&\frac{(D+F)^2(\mu_D-\mu_F)}{192\pi^3f_\pi^2}I_{2\pi}^{NN}
-\frac{1}{192\pi^3f_\pi^2}\left[(D-F)^2(2\mu_F+\mu_D)
I_{2K}^{N\Sigma}-(\frac D3+F)^2\mu_DI_{2K}^{N\Lambda}
\right. \\
&& \left.
-(D-F)(\frac{2D}3+2F)\mu_DI_{5K}^{N\Lambda\Sigma}\right]
-\frac{(\frac
D3-F)^2(\mu_D+3\mu_F)}{192\pi^3f_\pi^2}I_{2\eta}^{NN},
\end{eqnarray}
\begin{eqnarray}\nonumber
G_M^{n(1c)}&=&-\frac{(D+F)^2\mu_F}{96\pi^3f_\pi^2}I_{2\pi}^{NN}
-\frac{1}{192\pi^3f_\pi^2}\left[(D-F)^2(\mu_D-2\mu_F)I_{2K}^{N\Sigma}-(\frac D3+F)^2
\mu_DI_{2K}^{N\Lambda}
\right. \\
&& \left.
+(\frac{2D}3+2F)(D-F)\mu_DI_{5K}^{N\Lambda\Sigma}\right]
+\frac{(\frac D3-F)^2\mu_D}{96\pi^3f_\pi^2} I_{2\eta}^{NN},
\end{eqnarray}
where
\begin{equation}
I_{2j}^{\alpha\beta}=\int d\overrightarrow{k}\frac{k^2 u(\overrightarrow{k})^2}
{\omega_j(\overrightarrow{k})(\omega_j(\overrightarrow{k})+\delta^{\alpha\beta})^2},
\end{equation}
\begin{equation}
I_{5j}^{\alpha\beta\gamma}=\int d\overrightarrow{k}\frac{k^2 u(\overrightarrow{k})^2}
{\omega_j(\overrightarrow{k})(\omega_j(\overrightarrow{k})+\delta^{\alpha\beta})
(\omega_j(\overrightarrow{k})+\delta^{\alpha\gamma}))}.
\end{equation}

The magnetic moments of baryons in the chiral limit, expressed in
terms of $\mu_D$ and $\mu_F$, are used in the one loop calculations.
However, we have taken the mass difference of the octet baryons into
account.  If the masses of the octet baryons are taken to be
degenerate, then the coefficients in front of the integrals will be the
same as in the paper of Ref.~\cite{Jenkins2}.

The contributions to the form factors of Fig.~1d are expressed as
\begin{equation}
G_M^{p(1d)}=\frac{5{\cal C}^2\mu_C}{162\pi^3f_\pi^2}I_{2\pi}^{N\Delta}
+\frac{5{\cal C}^2\mu_C}{1296\pi^3f_\pi^2}I_{2K}^{N\Sigma^\ast},
\end{equation}
\begin{equation}
G_M^{n(1d)}=-\frac{5{\cal C}^2\mu_C}{648\pi^3f_\pi^2}I_{2\pi}^{N\Delta}
-\frac{5{\cal C}^2\mu_C}{1296\pi^3f_\pi^2}I_{2K}^{N\Sigma^\ast}.
\end{equation}
Fig.~1e and Fig.~1f give the following contributions to the form
factors:
\begin{equation}
G_M^{p(1e+1f)}=\frac{(D+F){\cal C}\mu_T}{108\pi^3f_\pi^2}I_{3\pi}^{N\Delta}
+\frac{5(D-F){\cal
C}\mu_T}{864\pi^3f_\pi^2}I_{5K}^{N\Sigma\Sigma^\ast}
+\frac{(D+3F){\cal
C}\mu_T}{864\pi^3f_\pi^2}I_{5K}^{N\Lambda\Sigma^\ast},
\end{equation}
\begin{equation}
G_M^{n(1e+1f)}=-\frac{(D+F){\cal C}\mu_T}{108\pi^3f_\pi^2}I_{3\pi}^{N\Delta}
+\frac{(D-F){\cal C}\mu_T}{864\pi^3f_\pi^2}I_{5K}^{N\Sigma\Sigma^\ast}
-\frac{(D+3F){\cal C}\mu_T}{864\pi^3f_\pi^2}I_{5K}^{N\Lambda\Sigma^\ast},
\end{equation}
where
\begin{equation}
I_{3j}^{\alpha\beta}=\int d\overrightarrow{k}\frac{k^2 u(\overrightarrow{k})^2}
{\omega_j(\overrightarrow{k})^2(\omega_j(\overrightarrow{k})+\delta^{\alpha\beta})}.
\end{equation}
Fig.~1g comes from the second order expansion of Lagrangian
(\ref{lomag}). The contributions to the magnetic form factors are
expressed as
\begin{equation}
G_M^{p(1g)}=-\frac{(\mu_D+\mu_F)}{32\pi^3f_\pi^2}I_{4\pi}
-\frac{\mu_F}{16\pi^3f_\pi^2}I_{4K},
\end{equation}
\begin{equation}
G_M^{n(1g)}=\frac{(\mu_D+\mu_F)}{32\pi^3f_\pi^2}I_{4\pi}
+\frac{(\mu_D-\mu_F)}{32\pi^3f_\pi^2}I_{4K},
\end{equation}
where
\begin{equation}
I_{4j}=\int d\overrightarrow{k}\frac{u(\overrightarrow{k})^2}
{\omega_j(\overrightarrow{k})}.
\end{equation}

The magnetic moment is defined as $\mu=G_M(Q^2=0)$. The total
nucleon magnetic moments can be written as
\begin{equation}\label{mup}
\mu_p(m_\pi^2)=a^p_0+a^p_2 m^2_\pi+a^p_4
m^4_\pi+\sum_{k=a}^g G_M^{p(1k)}(Q^2=0,m_\pi^2),
\end{equation}
\begin{equation}\label{mun}
\mu_n(m_\pi^2)=a^n_0+a^n_2 m^2_\pi+a^n_4
m^4_\pi+\sum_{k=a}^g G_M^{n(1k)}(Q^2=0,m_\pi^2),
\end{equation}
where $a^N_0$ ($N=n,p$) is expressed as
\begin{equation}
a^N_0=c^N_0-\sum_{k=a}^g G_M^{N(1k)}(Q^2=0,m_\pi^2=0).
\end{equation}
$c^p_0$ ($c^n_0$) is $\mu_p$ ($\mu_n$) which is related to $\mu_D$
and $\mu_F$ via Eq.~(\ref{treemag}). The residual series parameters,
$a_i$, are determined by the best fit of the lattice data.

\begin{table}
\caption{Residual series coefficients and nucleon form factors at zero
  momentum and 0.23 GeV$^2$.  The first two rows provide standard
  results whereas the final row includes the effect of the pion form
  factor.}
\begin{ruledtabular}
\begin{tabular}{ccccccccc}
$Q^2$ & $a_0^p$ & $a_0^n$ & $a_2^p$ (GeV$^{-2}$)& $a_2^n$
(GeV$^{-2}$) &$a_4^p$ (GeV$^{-4}$) & $a_4^n$ (GeV$^{-4}$) &
$G_M^p$ & $G_M^n$ \\ \hline
0 & 2.554 & $-$1.506 & $-$1.135 & 0.420 & 0.446 & $-$0.090 & 2.73 $\pm$ 0.20 & $-$1.84 $\pm$ 0.19 \\
0.23 & 1.617 & $-$0.932 & $-$0.411 & 0.070 & 0.144 & 0.031 & 1.70 $\pm$ 0.12 & $-$1.10 $\pm$ 0.11\\
0.23 & 1.652 & $-$0.968 & $-$0.499 & 0.159 & 0.201 & $-$0.027 & 1.65 $\pm$ 0.10 & $-$1.06 $\pm$ 0.09\\
\end{tabular}
\end{ruledtabular}
\end{table}

\section{Numerical results}

In the numerical calculations, the parameters are chosen as $D=0.76$
and $F=0.50$ ($g_A=D+F=1.26$). The coupling constant ${\cal C}$ is
chosen to be $-1.2$ which is the same as Ref.\ \cite{Jenkins2}.  The
renormalisation form factor $u(k)$ can be monopole, dipole or Gaussian
functions which give similar results \cite{Young2}.  In our
calculations, the dipole function is used:
\begin{equation}
u(k)=\frac1{(1+k^2/\Lambda^2)^2},
\end{equation}
with $\Lambda = 0.8$ GeV.  The coefficients $a^p_0$, $a^n_0$, $a^p_2$,
$a^n_2$, $a^p_4$ and $a^n_4$ in Eqs. (\ref{mup}) and (\ref{mun}) are
constrained by the quenched lattice data at large pion mass
($m_\pi>500$ MeV) where the quenched and physical values of the
magnetic moments are expected to be close to each other
\cite{Leinweber0,Young:2004tb}.

The $K$- and $\eta$-meson masses have relationships with the pion
mass according to
\begin{equation}
m_K^2=\frac12 m_\pi^2+m_K^2|_{\rm phy}-\frac12 m_\pi^2|_{\rm phy},
\end{equation}
\begin{equation}
m_\eta^2=\frac13 m_\pi^2+m_\eta^2|_{\rm phy}-\frac13 m_\pi^2|_{\rm
  phy}\, ,
\end{equation}
and enable a direct relationship between the meson dressings of the
nucleon magnetic moments and the pion mass.

We begin by considering nucleon form factor results from the CSSM
Lattice Collaboration \cite{Boinepalli:2006xd}.  The proton magnetic
moment $\mu_p$ versus $m_\pi^2$ is shown in Fig.~2.  Here, the last
five lattice points at larger $m_\pi^2$ are used in the fit to avoid
quenched chiral artifacts.  The lines with label a and b+c+d+e+f+g
correspond to the contributions of Fig.~1a and sum of the other
diagrams, respectively. The residual series contribution, {\it i.e.}
the contribution from $a_0+a_2 m_\pi^2+a_4 m_\pi^4$ is also shown in
the figure and labeled by ``tree''.  The near linear behavior of the
residual series is a reflection of the excellent convergence of the
residual expansion.

\begin{center}
\begin{figure}[tbp]
\includegraphics[scale=0.9]{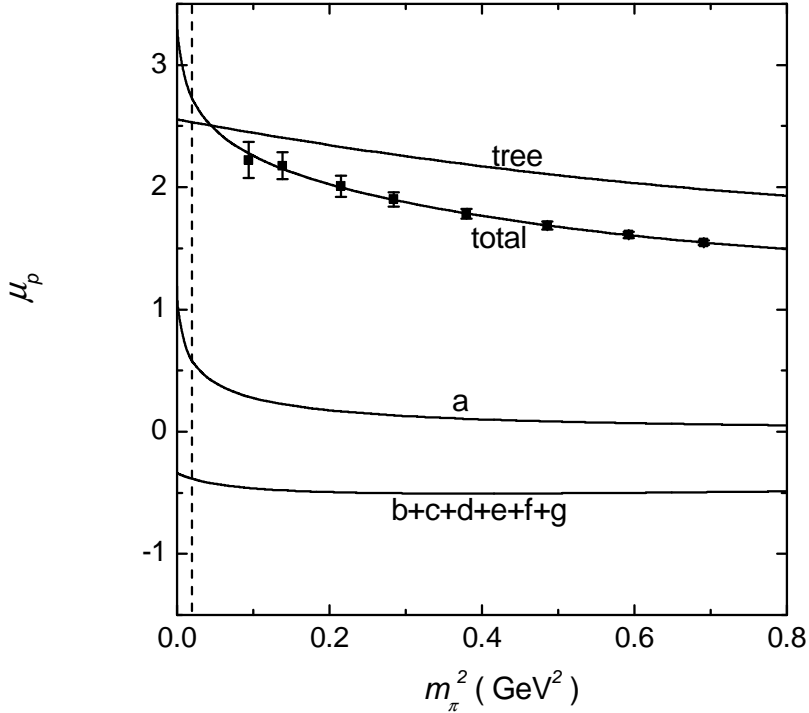}
\caption{The proton magnetic moment \cite{Boinepalli:2006xd} versus
squared pion mass.  The lines with label a and b+c+d+e+f+g correspond
to the contributions of Fig.~1a and the sum of the other diagrams,
respectively.}
\end{figure}
\end{center}

The leading diagram (Fig.~1a) gives the dominant chiral behavior of
the magnetic moment.  At small pion mass, the proton magnetic moment
decreases quickly with the increasing pion mass.  At larger pion mass,
the proton magnetic moment changes smoothly. At the physical point,
$m_\pi=0.139$ GeV, the proton magnetic moment is $2.73\ \mu_N$, close
to the experimental value, $2.79\ \mu_N$.  We emphasize again that the
chiral curvature is dominated by Fig.~1a.

The neutron magnetic moment, $\mu_n$, is studied in the same way.
$\mu_n$ versus $m_\pi^2$ is shown in Fig.~3.  Again, only the five
lattice points at larger pion mass are used in the fit to avoid
quenched chiral artifacts.  Similar to what was found in proton
case, the leading diagram gives the dominant chiral curvature. The
neutron magnetic moment increases quickly as ones moves from the
chiral limit and becomes smooth at large pion mass. At the physical
point, the neutron magnetic moment is $-1.84\ \mu_N$, compares
favorably with the experimental value of $-1.91\mu_N$.
For both the proton and neutron, the rapid variation of the magnetic moments
at $m_\pi=0$ may reflect the fact that the nucleon magnetic radii diverge
in the chiral limit.
\begin{center}
\begin{figure}[tbp]
\includegraphics[scale=0.9]{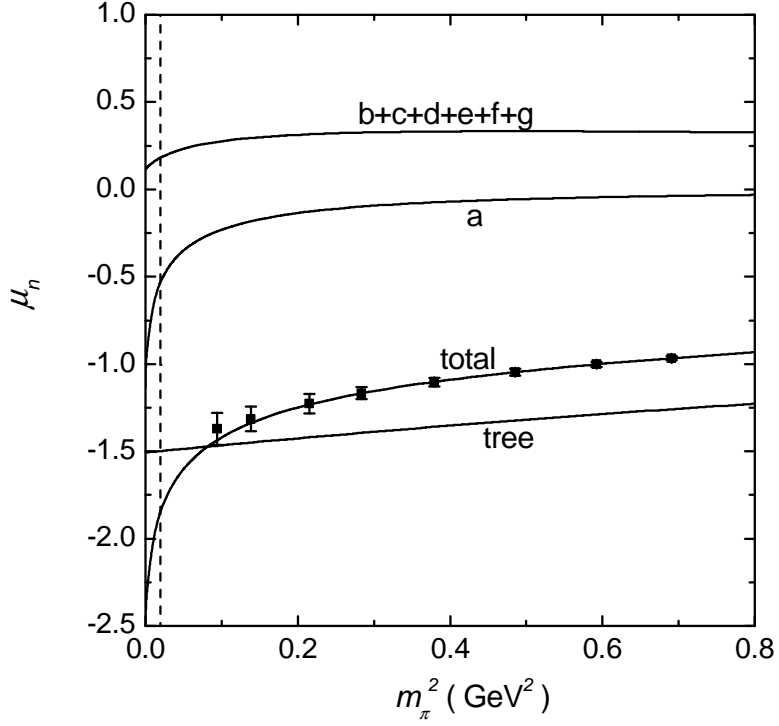}
\caption{The neutron magnetic moment \cite{Boinepalli:2006xd} versus
squared pion mass. The lines with label a and b+c+d+e+f+g correspond
to the contributions of Fig.~1a and the sum of the other diagrams,
respectively.}
\end{figure}
\end{center}

In the above numerical calculations, we selected $\Lambda$ to be 0.8
GeV.  In
Fig.~4 we show the nucleon magnetic moments versus $\Lambda$.  The
proton magnetic moment and the absolute value of neutron
magnetic moment increase almost linearly with increasing $\Lambda$.
In the range $0.6<\Lambda<1.0$ GeV, the proton (neutron) magnetic
moment varies from $2.55\ \mu_N$ ($-1.66\ \mu_N$) to $2.90\ \mu_N$
($-2.02\ \mu_N$). When $\Lambda$ is around 0.8 GeV, both the proton and
neutron magnetic moments are in good agreement with the experimental
values.

Since the extrapolated values are $\Lambda$ dependent (an indication
that the fits lie outside the power-counting regime), the uncertainty
of $\Lambda$ will result in an additional source of error in the final
result.  Through a consideration of optimizing the convergence
properties of the finite-range regularised expansion, we include the
variation of $\Lambda$ in the range 0.8 $\pm$ 0.2 GeV and add this
uncertainty to the statistical uncertainties in quadrature.  The
extrapolated magnetic moments with corresponding error bars are listed
in table I.

In the chiral limit, the magnetic moments, $c_0^p$ and $c_0^n$, are
3.41 and -2.53. These two values are close to the corresponding
ones used in normal chiral perturbation theory. For example, in
Ref.~\cite{Fuchs:2003ir}, the corresponding values are 3.38 and
-2.66. For the higher order terms, our low energy constants are
much smaller, resulting in more convergent behavior. For example,
our $a_2^p$ and $a_2^n$ are $-1.14$ and 0.42 which are much smaller
than the corresponding values $-6.80$ and 8.75 in Ref.~\cite{Fuchs:2003ir}.
\begin{figure}[tbp]
\begin{center}
\includegraphics[scale=0.9]{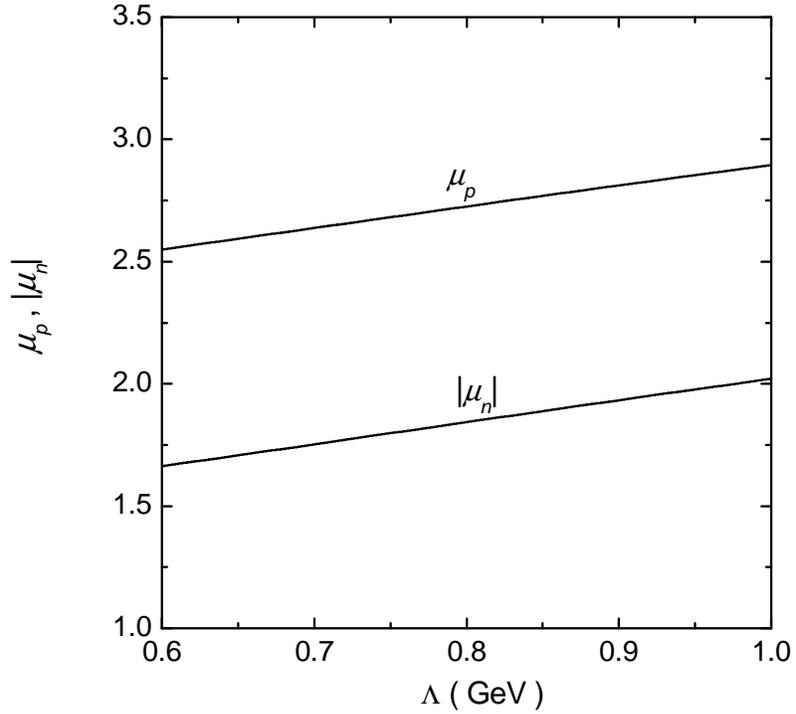}
\caption{The proton and neutron magnetic moments versus the regulator
  scale $\Lambda$.}
\end{center}
\end{figure}

We now proceed to extrapolate the nucleon magnetic form factors at
finite $Q^2$.  At finite momentum transfer, we choose not to express
the magnetic form factors in terms of $Q^2$ and $Q^4$ as we did for
the $m_\pi$ dependence of $\mu_N$.  This is because the momentum
dependence of form factors is close to the following assumption
$G_M^N(Q^2)=\mu_N/(1+Q^2/0.71$GeV$^2)^2$.  The high order terms in a
$Q^2$ expansion are important and the truncation of momentum to some
order, say fourth order, is not a good approximation.

In our calculation, the same formulas as Eqs. (\ref{mup}) and
(\ref{mun}) are used for the extrapolation of the magnetic form
factors at each fixed finite value of $Q^2$.  The $Q^2$ dependence of
nucleon magnetic form factors at tree level is included in the
parameters $a_0$, $a_2$ and $a_4$, as these parameters are constrained
by the lattice results at finite $Q^2$.

In addition to the CSSM Lattice collaboration results
\cite{Boinepalli:2006xd} considered thus far, we also consider QCDSF
lattice results at finite $Q^2$ \cite{QCDSF}.  Large statistical
uncertainties encountered at large $Q^2$ prevent one from constraining
the $m_\pi^{4}$ term and therefore we fit the QCDSF data using a
residual series expansion up to and including order $m_\pi^{2}$ only.
The coefficients together with the form factors at finite $Q^2$ are
obtained by fitting the lattice results and are listed in Tables I and
II.  From the tables, one can see that $a_0^p$ decreases with
increasing momentum, while $a_0^n$ increases with momentum. $a_2^N$
(N=p,n) are small indicating good convergence of the expansion.

\begin{table}
\caption{Residual series coefficients and nucleon form factors at
  various $Q^2$.  The first four lines report standard results while
  the latter four lines report results including the effect of a pion
  form factor.}
\begin{ruledtabular}
\begin{tabular}{ccccccc}
$Q^2$ (GeV$^2$) & $a_0^p$ & $a_0^n$ & $a_2^p$ (GeV$^{-2}$)& $a_2^n$
(GeV$^{-2}$) & $G_M^p$ & $G_M^n$ \\
\hline
0.557 & 1.042 & $-$0.638 & $-$0.024  & $-$0.00 & 1.07 $\pm$ 0.17 & $-$0.71 $\pm$ 0.14 \\
1.08  & 0.609 & $-$0.337 &    0.015  & $-$0.04 & 0.61 $\pm$ 0.13 & $-$0.37 $\pm$ 0.13 \\
1.14  & 0.598 & $-$0.348 &    0.052  & $-$0.04 & 0.59 $\pm$ 0.11 & $-$0.37 $\pm$ 0.09 \\
2.28  & 0.293 & $-$0.178 &    0.035  & $-$0.03 & 0.28 $\pm$ 0.09 & $-$0.18 $\pm$ 0.05 \\
\hline
0.557 & 1.051 & $-$0.650 & $-$0.033  &    0.01 & 1.01 $\pm$ 0.15 & $-$0.66 $\pm$ 0.10 \\
1.08  & 0.620 & $-$0.349 &    0.008  & $-$0.03 & 0.58 $\pm$ 0.12 & $-$0.34 $\pm$ 0.12 \\
1.14  & 0.610 & $-$0.360 &    0.044  & $-$0.04 & 0.57 $\pm$ 0.10 & $-$0.35 $\pm$ 0.07 \\
2.28  & 0.300 & $-$0.185 &    0.032  & $-$0.02 & 0.27 $\pm$ 0.09 & $-$0.17 $\pm$ 0.05 \\
\end{tabular}
\end{ruledtabular}
\end{table}

We plot the $m_\pi$ dependence of proton and neutron magnetic form
factors at $Q^2=0.23$ GeV$^2$ in Figs.~5 and 6 respectively.  At small
pion mass, the proton and neutron magnetic form factors do not change
as quickly as in the case of zero momentum.  However, the diagram of
Fig.~1a still gives the dominant contribution to the curvature when
the pion mass is small.  From the figures, one can see that the
expansion in powers of $m_\pi$ shows good convergence.  At the
physical pion mass, $G_M^p(0.23$ GeV$^2)=(1.70$ $\pm$ 0.12) $\mu_N$ and
$G_M^n(0.23$ GeV$^2)=-1.10$ $\pm$ 0.11) $\mu_N$, which are both
reasonable compared with experiment.

\begin{figure}[tbp]
\begin{center}
\includegraphics[scale=0.9]{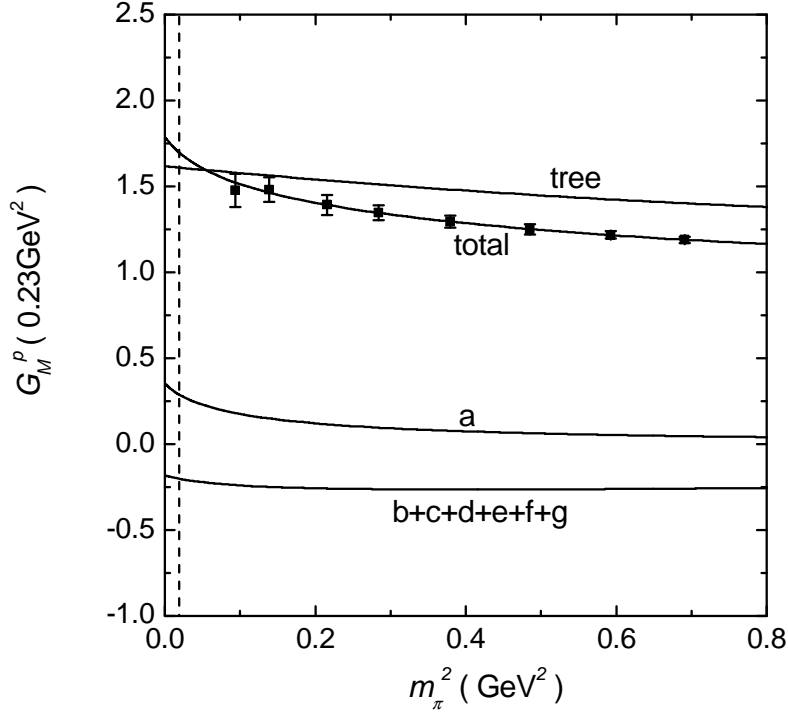}
\caption{The proton magnetic form factor \cite{Boinepalli:2006xd} at
$Q^2=0.23$ GeV$^2$ versus the squared pion mass.  The lines with label
a and b+c+d+e+f+g correspond to the contributions of Fig.~1a and the
sum of the other diagrams respectively.}
\end{center}
\end{figure}

\begin{figure}[tbp]
\begin{center}
\includegraphics[scale=0.9]{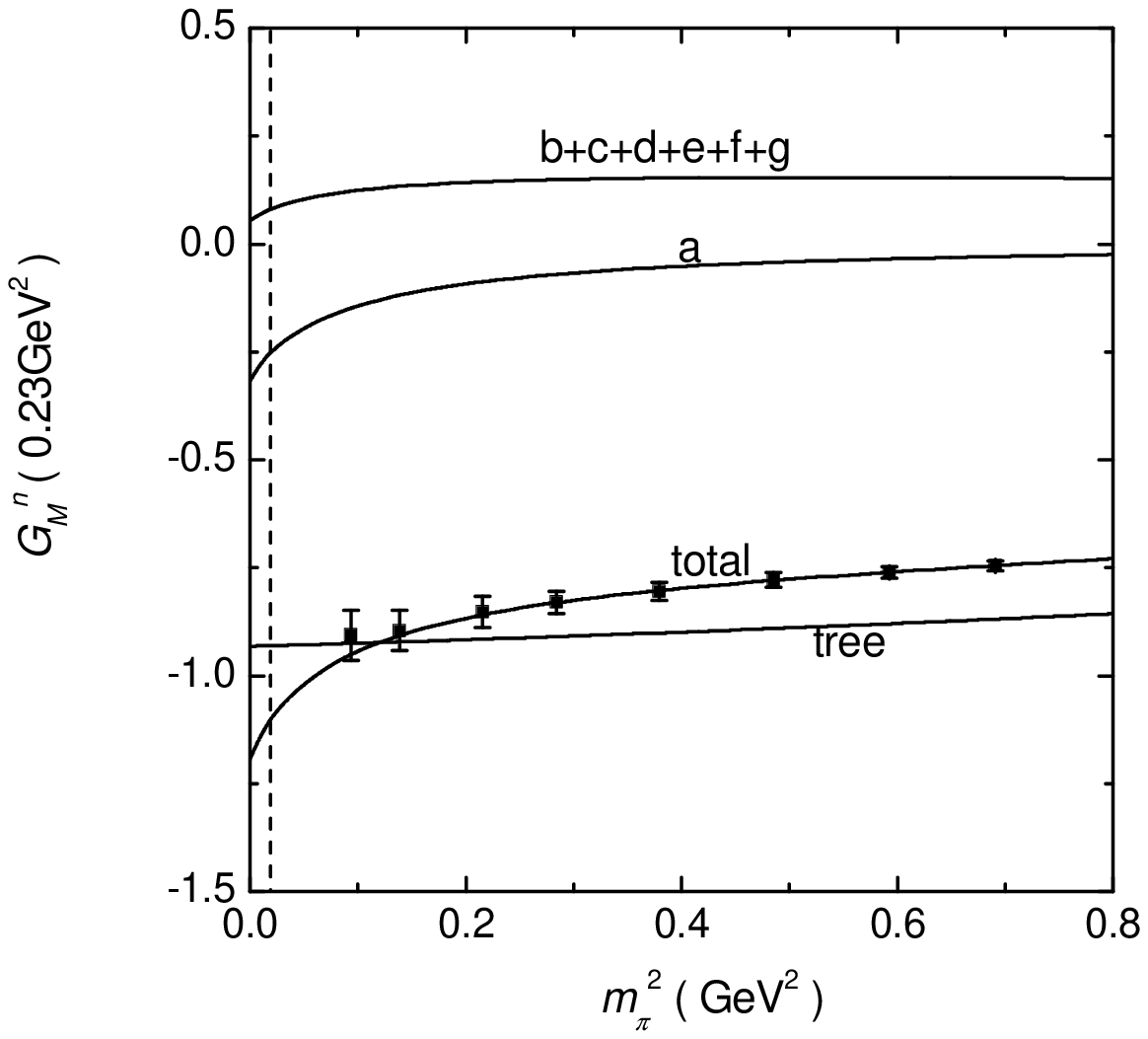}
\caption{The neutron magnetic form factor \cite{Boinepalli:2006xd} at
$Q^2=0.23$ GeV$^2$ versus squared pion mass. The lines with label a
and b+c+d+e+f+g correspond to the contributions of Fig.\ 1a and the
sum of the other diagrams respectively.}
\end{center}
\end{figure}

In Figs.~7 and 8, we plot the proton and neutron magnetic form factors
respectively.  Results at $Q^2=0.557$, 1.08, 1.14 and 2.28 GeV$^2$
from the QCDSF collaboration \cite{QCDSF} are considered.  From the
figures, one can see that the lattice data do not vary smoothly as a
function of the pion mass due to large statistical errors.  As a
consequence, the extrapolated magnetic form factors at the physical
pion mass have relatively large error bars.  Accurate lattice results
are needed to better constrain the chiral expansion parameters and
allow one to consider an $m_\pi^4$ term in the residual expansion.

\begin{center}
\begin{figure}[tbp]
\includegraphics[scale=0.9]{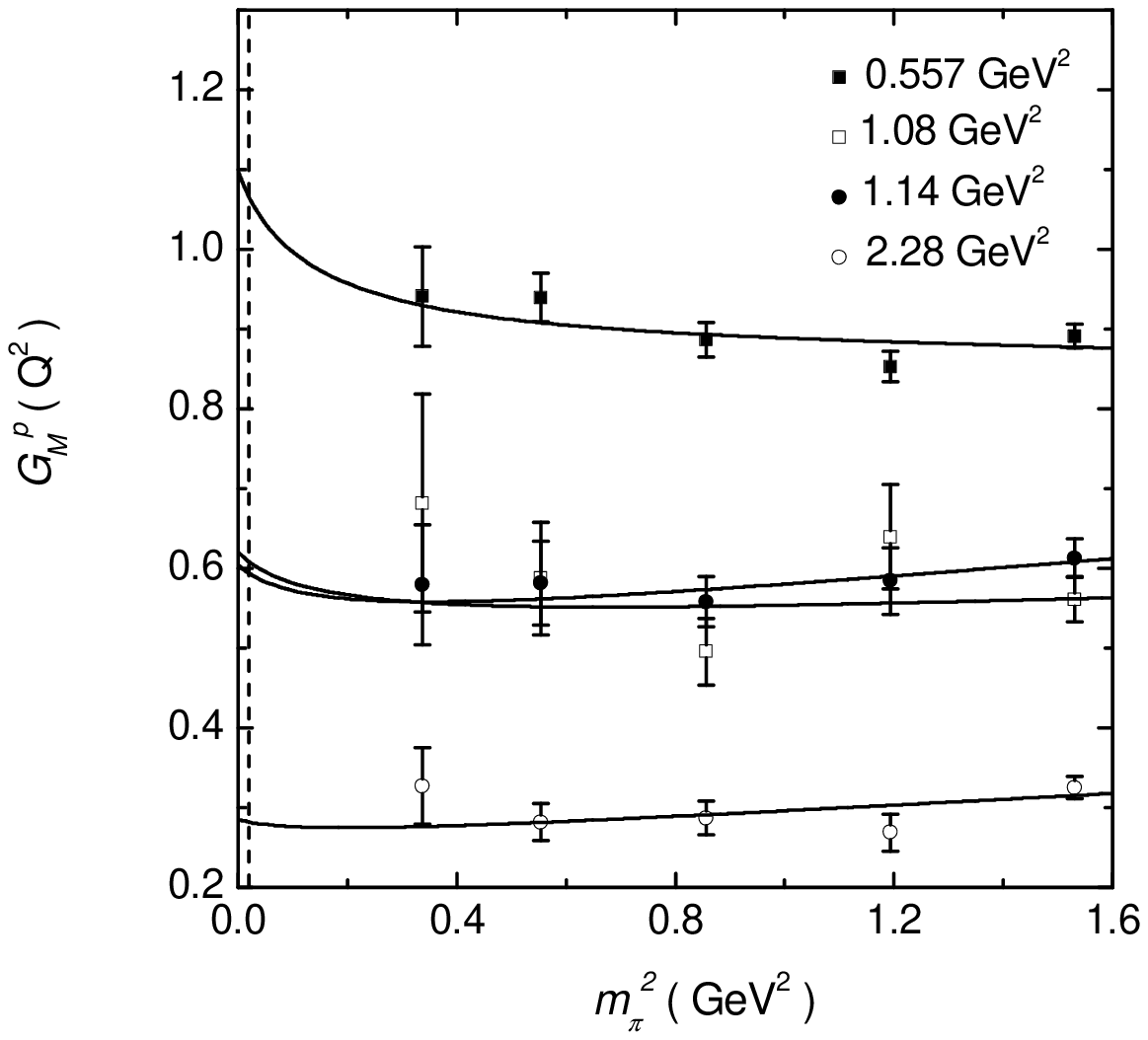}
\caption{The proton magnetic form factors \cite{QCDSF} at $Q^2=0.557$,
  1.08, 1.14 and 2.28 GeV$^2$ versus pion mass.}
\end{figure}
\end{center}

\begin{center}
\begin{figure}[tbp]
\includegraphics[scale=0.9]{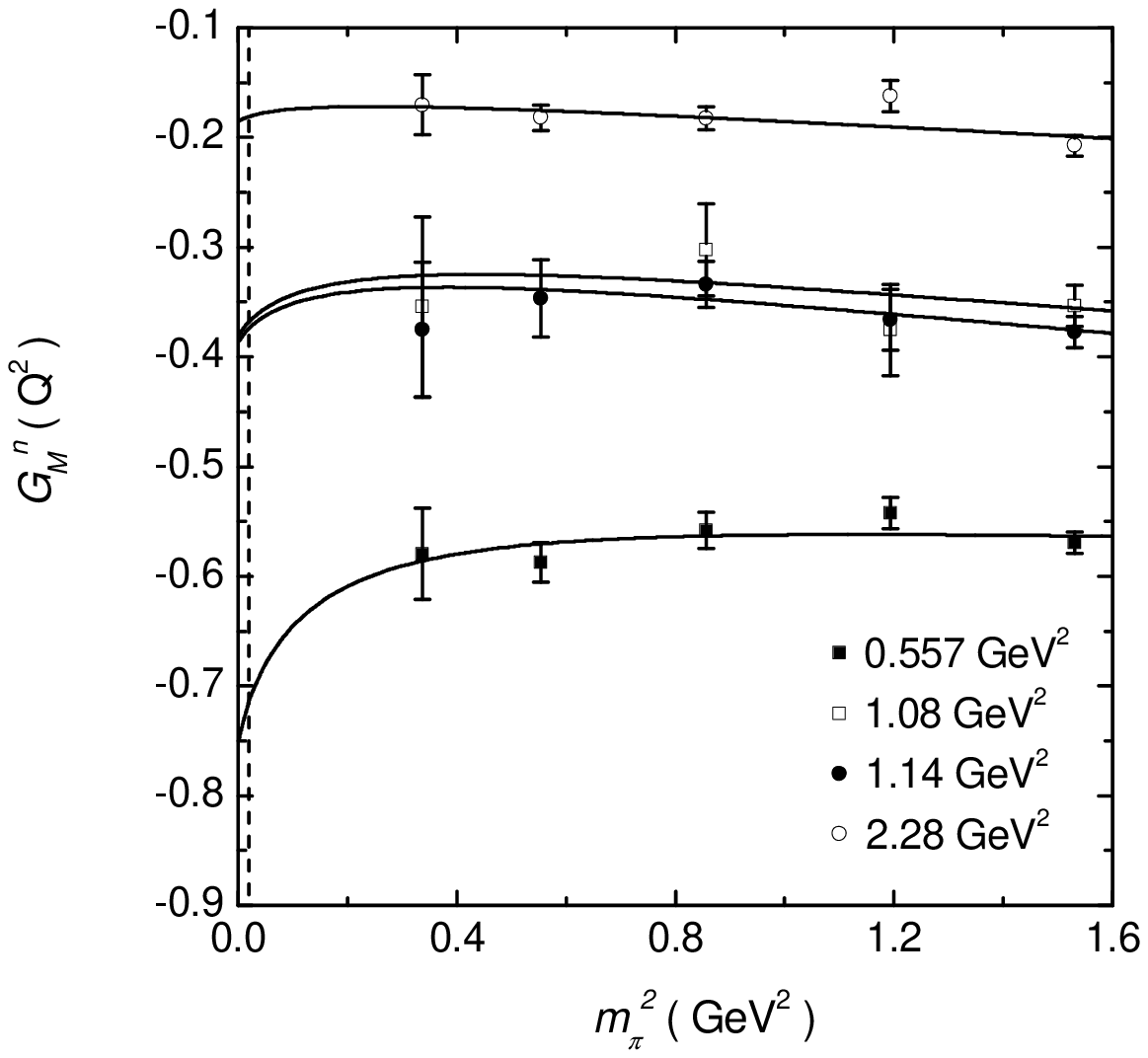}
\caption{The neutron magnetic form factors \cite{QCDSF} at
  $Q^2=0.557$, 1.08, 1.14 and 2.28 GeV$^2$ versus pion mass.}
\end{figure}
\end{center}

To this point, we have not considered the possibility of an important
role for the pion form factor in the calculation.  We know that at
large $Q^2$, the pion form factor is much less than one and this will
affect the meson cloud contribution to nucleon magnetic form factors.
We list the pion electromagnetic form factor $F_\pi$ in Table III as
provided in Ref.\ \cite{Melo}.

With these pion form factors, we repeat the chiral fit of the nucleon
magnetic form factors.  As an example, in Fig.~9, we show the result
obtained for the proton magnetic form factor at $Q^2$=0.23 GeV$^2$.
The dashed and solid lines are for the results with and without the
pion form factor consideration, respectively.  When the pion form
factor is included, the leading diagram Fig.~1a provides less
curvature.  As a result, the total $G_M^P$ decreases from 1.70 $\pm$
0.12 $\mu_N$ to 1.65 $\pm$ 0.10 $\mu_N$.  For the neutron, $G_M^n$
increases from $-1.10$ $\pm$ 0.11 $\mu_N$ to $-1.06$ $\pm$ 0.09
$\mu_N$.  Though the pion form factor changes significantly at finite
momentum, it does not affect the nucleon magnetic form factors very
much.  At large $Q^2$, the pion form factor has a negligible effect on
the nucleon form factors because the loop contribution itself is
already very small.

\begin{table}[b]
\caption{Pion electromagnetic form factor, $F_\pi$, at various $Q^2$.}
\begin{ruledtabular}
\begin{tabular}{cccccc}
$Q^2$ (GeV$^2$) & 0.23 & 0.557 & 1.08 & 1.14 & 2.28 \\
\hline
$F_\pi$ & 0.70 & 0.50 & 0.31 & 0.29 & 0.18\\
\end{tabular}
\end{ruledtabular}
\end{table}

\begin{figure}[tbp]
\begin{center}
\includegraphics[scale=0.9]{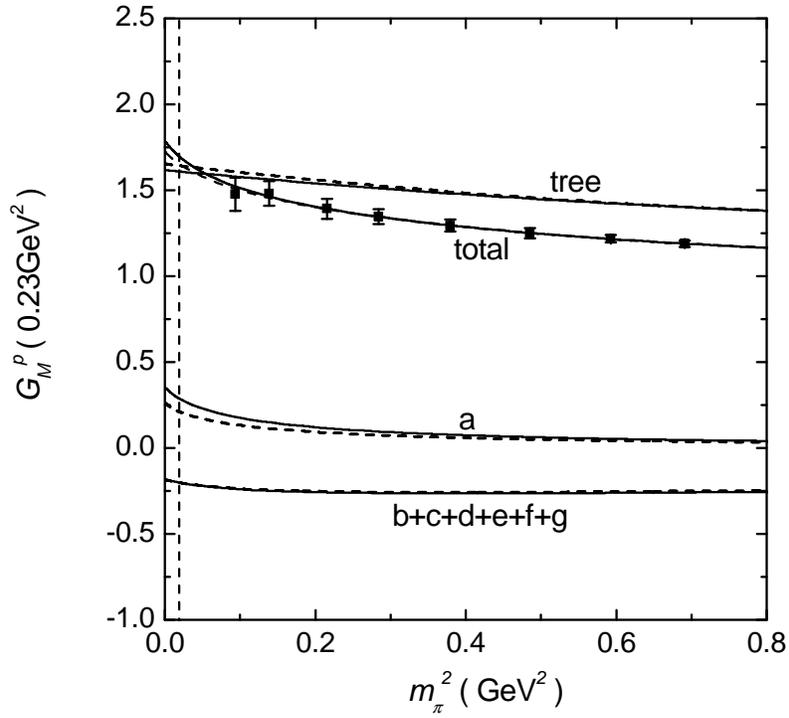}
\caption{The proton magnetic form factor \cite{Boinepalli:2006xd} at
  $Q^2=0.23$ GeV$^2$ versus pion mass.  The dashed and solid lines are
  for the results with and without the pion form factor,
  respectively.}
\end{center}
\end{figure}

In Fig.~10, we show the extrapolated proton and neutron magnetic form
factors versus $Q^2$ at the physical pion mass with the corresponding
error bars. The hollow and solid square points are for the fits with
and without the pion form factor, respectively.

The solid lines in Fig.~10 are the empirical parametrization
$G_M^N(Q^2)=\mu_N/(1+Q^2/0.71$GeV$^2)^2$.  At both zero momentum and
$Q^2=0.23$ GeV$^2$, the extrapolated CSSM results are in good
agreement with the experimental data.  For the other values of $Q^2$
from the QCDSF collaboration, the extrapolated nucleon magnetic form
factors are in reasonable agreement with the empirical
parameterization. For the neutron the agreement is quite reasonable,
while for the proton, although the extrapolation is consistently
within one standard deviation of the empirical curve, the
extrapolated values do appear to be systematically a little high.

We should mention that we use just two parameters, $a_0$ and $a_2$,
to fit the lattice data (at each value of $Q^2$). We note that it
would be very helpful to have more accurate lattice data over a
range of lattice spacings and volumes in order to extrapolate to the
infinite volume continuum limit and to be able to incorporate an
$a_4$ term. One would also prefer to work with full QCD data rather
than quenched data. Until these conditions are satisfied it is a
little early to draw strong conclusions about the validity of the
extrapolation in pion mass from a comparison with experimental data
for the form factors. Indeed, that the current results lie within
one standard deviation of the data at all values of $Q^2$ is really
a very positive result at the present stage. We do emphasis that our
results are based on the lowest order Lagrangian at one loop level
in heavy baryon approximation. That must breakdown at high momentum
transfer, however, that is precisely where, in the FRR treatment,
the loops become naturally small -- for the clear physical reason
that high pion momenta (and high pion mass) are suppressed by the
finite size of the source. The extrapolation error arising from the
uncertainty in $\Lambda$ included above is small when the momentum
is high.  Again, this is because at high momentum transfer, the loop
contribution to the total magnetic form factor is small.

\begin{figure}[tbp]
\begin{center}
\includegraphics[scale=0.9]{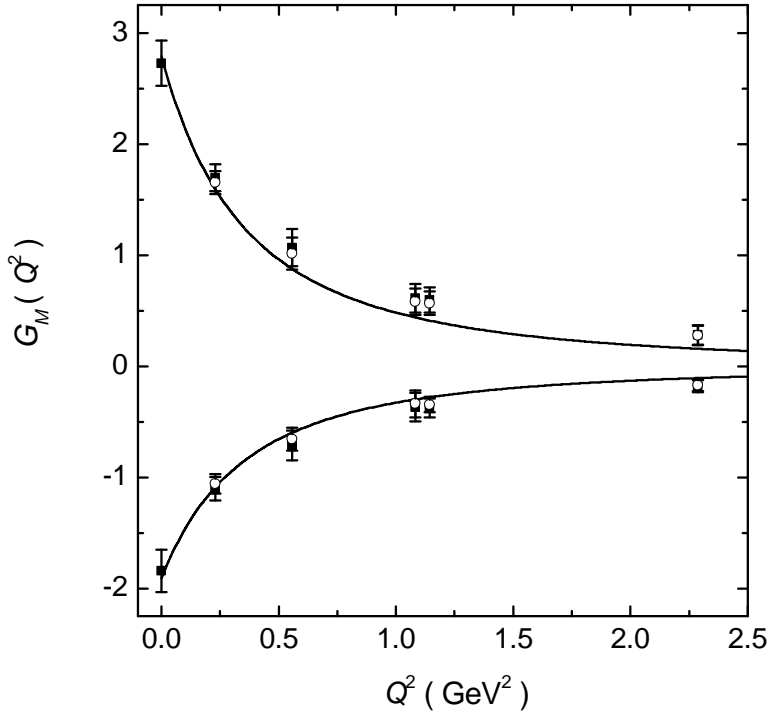}
\caption{The proton and neutron magnetic form factors at physical pion
  mass versus the momentum transfer, $Q^2$.  The hollow and solid
  square points are for the fits with and without the
  inclusion of the form factor of the pion, respectively.}
\end{center}
\end{figure}

\section{summary}

We extrapolated state of the art lattice results for nucleon magnetic
form factors in an extension of heavy baryon chiral perturbation theory.  All one-loop
graphs are considered at arbitrary momentum transfer and all
octet and decuplet baryons are included in the intermediate states.

Finite-range regularisation is used in the one loop calculation to
improve the convergence of the chiral expansion.  The residual series
coefficients $a_0^p$($a_0^n$), $a_2^p$ ($a_2^n$) and $a_4^p$ ($a_4^n$)
are obtained by fitting the lattice results at $m_\pi>0.5$ GeV, where
quenched artifacts are anticipated to be small.

The leading non-analytic diagram provides the dominant curvature for
the $m_\pi$ dependence of magnetic moments.  The sum of higher-order
one-loop terms provide only a small correction to this curvature.  The
one-loop contributions show that the proton (neutron) magnetic moment
decreases (increases) quickly with increasing pion mass in the small
$m_\pi$ region. At larger pion masses, their contributions change
slowly and smoothly.

The magnetic form factors are also studied at large $Q^2$ where
chiral nonanalytic behavior is suppressed.  Here, the importance of
the pion form factor is also examined. For $Q^2=0.23$ GeV$^2$,
$G_M^p= 1.70\pm 0.12\ \mu_N$ and $G_M^n= -1.10\pm 0.11\ \mu_N$. Upon
including the pion form factor, these values will change to $1.65\pm
0.10\ \mu_N$ and $-1.06 \pm 0.09\ \mu_N$ indicating the effect is
subtle.  Although the pion form factor decreases quickly with
increasing momentum, its effect on nucleon form factors is not
significant, as the loop integrals themselves are already small.

The chirally extrapolated results of Fig.~10 compare favorably with
experiment and demonstrate the utility and accuracy of the chiral
extrapolation methods presented herein. With the more accurate
lattice data and treatment, it is of interest to see whether the
mismatch at large $Q^2$ will disappear or not.

\section*{Acknowledgements}

P.W. thanks the Theory Group at Jefferson Lab for their kind
hospitality. This work was supported by the Australian Research
Council and by DOE contract DOE-AC05-06OR23177, under which Jefferson
Science Associates operates Jefferson Lab.

\end{document}